# Solutions of the *D*-dimensional Schrödinger equation with the Hyperbolic Pöschl–Teller potential plus modified ring shaped term

**Ibsal A. Assi** [1], **Akpan N. Ikot**[2*] **and E.O. Chukwuocha**[2]

[1] *Physics Department, King Fahd University of Petroleum & Minerals, Dhahran 31261, Saudi Arabia*

[2]*Department of Physics, University of Port Harcourt, P.M.B 5323, Choba, Port Harcourt, Nigeria*

*email:ndemikotphysics@gmail.com

**Abstract**: In this paper, we solve the *D*-dimensional Schrödinger equation with hyperbolic Poschl-Teller potential plus a generalized ring-shaped potential. After the separation of variable in the hyperspherical coordinate. We used Nikiforov-Uvarov (NU) method to solve the resulting radial equation and obtain explicitly the energy level and the corresponding wave function in closed form. The solutions to the angular part are solved using the NU approach as well.

## 1. Introduction

The non-central potentials in recent times have been an active field of research in physics and quantum chemistry [1-3]. For instance, the occurrence of accidental degeneracy and hidden symmetry in the non-central potentials and its application in quantum chemistry and nuclear physics are used to describe ring-shaped molecules like benzene and the interaction between deformed pair of nuclei [4-5]. It is known that these accidental degeneracy occurring in the ring shaped was explain by constructing an SU (2) algebra [6]. Owing to these applications many authors have investigated a number of real physical problems on non-spherical oscillator [7], ring-shaped oscillator (RSO) [8] and ring shaped non-spherical oscillator [9]. Berdemir [10] had shown that either Coulomb or harmonic oscillator will give a better approximation for understanding the spectroscopy and structure of diatomic molecules in the ground electronic state. Other applications of the ring shaped potential can be found in ring shaped organic molecules like cyclic polyenes and benzene [11-12].

On the other hand, Chen and Dong studied the Schrödinger equation with a new ring shaped potential [13]. Cheng and Dai investigated modified Kratzer potential plus the new ring shaped potential using Nikiforov-Uvarov method [14]. Recently, Ikot *et al* [15-17] investigated the Schrödinger equation with Hulthen plus a new ring shape potential [13],Non-spherical harmonic and Coulomb potential[16] and pseudo-Coulomb potential in the cosmic string  space-time[17].Many authors have used different methods to obtain exact solutions of the wave equation such as the methods of Supersymmetric Quantum Mechanics (SUSY-QM) [18-20], the Tridiagonal Representation Approach (TRA) [21-24], Nikiforov-Uvarov (NU) [25-29] among other methods

Motivated by the recent studies of the ring shaped-like potential [30-33], we proposed a novel hyperbolical Poschl Teller potential plus generalized ring shaped potential of the form,

$$V(r,\theta) = A\tanh^2(\lambda r) + \frac{B}{\tanh^2(\lambda r)} + \frac{\gamma\cot^2\theta + \zeta\cot\theta\csc\theta + \kappa\csc^2\theta}{r^2}, \quad (1.1)$$

where $\lambda$ is the screening parameter, $A$, $B$, $\gamma$, $\zeta$, and $\kappa$ are real potential parameters. As a special case when $\lambda r \to 0$ with $A \to \frac{m\omega^2}{2\lambda^2} - \frac{\hbar^2\alpha}{30m}\lambda^2$, $B \to \frac{\hbar^2\alpha}{2m}\lambda^2$, and $E \to E + \frac{2B}{3}$, the potential of equation (1.1) turns to Non-spherical harmonic oscillator plus generalized ring shaped potential

$$V(r,\theta) = \frac{1}{2}m\omega^2 r^2 + \frac{\hbar^2\alpha}{2mr^2} + \frac{\gamma\cot^2\theta + \zeta\cot\theta\csc\theta + \kappa\csc^2\theta}{r^2}, \quad (1.2)$$

## 2. *D*-Dimensional Schrodinger equation in Hyperspherical Coordinates

The *D*-dimensional Schrodinger equation is given below [34-35]

$$\left\{\nabla_D^2 + \frac{2\mu}{\hbar^2}[E - U]\right\}\Psi_{l_1,l_2,...,l_{D-2}}^{l=l_{D-1}}(\vec{X}) = 0, \quad (2.1)$$

where $\mu$ is the effective mass of two interacting particles, $\hbar$ is Planck's constant, $E$ is the energy eigenvalue, $U$ is the potential energy function, $\vec{X} = (r,\theta_1,\theta_2,....,\theta_{D-1})^T$ is the position vector in *D*-dimensions, where $\{\vec{\theta}\} = \{\theta_1,\theta_2,......,\theta_{D-1}\}$ is the angular position vector written in terms of Hyperspherical coordinates [36-37], and $\nabla_D^2$ is the *D*-dimensional Laplacian operator given in Appendix B.

The solvable potentials which allows separation of variable in (2.1) must be of the form:

$$U(\vec{x}) = V_1(r) + \frac{V_2(\theta_{D-1})}{r^2}, \quad (2.2)$$

. The separable wavefunction take the following form:

$$\Psi_{l_1,l_2,...,l_{D-2}}^{l=l_{D-1}}(\vec{X}) = r^{-(D-1)/2} g(r) Y_{l_1,l_2,...,l_{D-2}}^{l=l_{D-1}}(\vec{\theta}), \quad (2.3)$$

Applying (2.3) to Eq. (2.1) with the use of Eq. (2.2), we obtain the following radial and angular wave equations

$$\left[\frac{d^2}{dr^2} - \frac{(D+2l-2)^2 - 1}{4r^2} + \frac{2\mu}{\hbar^2}[E - V_1(r)]\right]g(r) = 0, \quad (2.4)$$

$$\left[\frac{d^2}{d\theta_j^2}+(j-1)\frac{\cos\theta_j}{\sin\theta_j}\frac{d}{d\theta_j}+\Lambda_j-\frac{\Lambda_{j-1}}{\sin^2\theta_j}\right]H(\theta_j)=0, \qquad (2.5)$$

$$\left[\frac{d^2}{d\theta_{D-1}^2}+(D-2)\frac{\cos\theta_{D-1}}{\sin\theta_{D-1}}\frac{d}{d\theta_{D-1}}+l(l+D-2)-\frac{\Lambda_{D-2}}{\sin^2\theta_{D-1}}+\frac{2\mu}{\hbar^2}V_2(\theta_{D-1})\right]H(\theta_{D-1})=0, \quad (2.6)$$

where $Y_{l_1,l_2,\ldots,l_{D-2}}^{l=l_{D-1}}(\vec{\theta})=\frac{1}{\sqrt{2\pi}}e^{\pm im\theta_1}\prod_{j=2}^{D-1}H(\theta_j)$, Eq. (2.6) holds for $j\in[2,D-2]$, with $D>3$, and $\Lambda_j=l_j(l_j+j-1)$. Solutions of (2.6) will not be affected by the presence of the proposed potential and thus it is common to different systems and it was done before using different approaches [38]. Consequently, we will only solve equations (2.4) and (2.6) using the method of Nikiforov-Uvarov [25,26].

## 3. Nikiforov-Uvarov Method

Many problems in physics leads to the following second order linear differential equation [25]:

$$\left[\frac{d^2}{dx^2}+\frac{\tilde{\tau}(x)}{\sigma(x)}\frac{d}{dx}+\frac{\tilde{\sigma}(x)}{\sigma^2(x)}\right]u(x)=0, \qquad (3.1)$$

where $\sigma(x)$, $\tilde{\sigma}(x)$ are polynomials of degree at most 2, and $\tilde{\tau}(x)$ is at most linear in $x$. Eq. (3.1) is sometimes called of hypergeometric type. Let's consider $u(x)=\phi(x)y(x)$, this will transform equation (3.1) to the following differential equation for $y(x)$:

$$\left[\frac{d^2}{dx^2}+\frac{\tau(x)}{\sigma(x)}\frac{d}{dx}+\frac{\bar{\sigma}(x)}{\sigma^2(x)}\right]y(x)=0, \qquad (3.2)$$

where we assumed the following conditions:

$$\frac{\phi'(x)}{\phi(x)}=\frac{\pi(x)}{\sigma(x)}, \qquad (3.3.\text{a})$$

$$\tau(x)=\tilde{\tau}(x)+2\pi(x), \qquad (3.3.\text{b})$$

$$\bar{\sigma}(x)=\tilde{\sigma}(x)+\pi^2(x)+\pi(x)\left[\tilde{\tau}(x)-\sigma'(x)\right]+\pi'(x)\sigma(x), \qquad (3.3.\text{c})$$

$$\pi(x)=\frac{\sigma'(x)-\tilde{\tau}(x)}{2}\pm\sqrt{\left(\frac{\sigma'(x)-\tilde{\tau}(x)}{2}\right)^2-\tilde{\sigma}(x)+k\sigma(x)}, \qquad (3.3.\text{e})$$

$$k=\eta-\pi'(x), \qquad (3.3.\text{f})$$

where $k$ and $\eta$ are constants chosen such that $\pi(x)$ is polynomial which at most linear in $x$ and $\bar{\sigma}(x) = \eta\sigma(x)$. This will transform Eq. (3.2) to the following:

$$\left[\sigma(x)\frac{d^2}{dx^2} + \tau(x)\frac{d}{dx} + \eta\right]y(x) = 0, \qquad (3.4)$$

where $\sigma(x)$ and $\tau(x)$ are polynomials of degrees 2 and 1, respectively. In this case, solutions to Eq. (3.4) are polynomials of degree $n$, $y(x) = y_n(x,\eta_n)$, where $\eta_n$ is given below:

$$\eta_n = -n\tau' - \frac{1}{2}n(n-1)\sigma'', \qquad (3.5)$$

Eq. (3.5) will be used to obtain the energy spectrum formula of the quantum mechanical system. We should point out here that the polynomial solutions to Eq. (3.4) for $\tau' < 0$ and $\tau = 0$ on the boundaries of the finite space (the latter case is omitted for infinite space), are the classical orthogonal polynomials. It is well known that each set of polynomials is associated with a weight function $\rho(x)$. For the polynomial solutions to Eq. (3.4), this function must be bounded on the domain of the system and must satisfy $(\sigma\rho)' = \tau\rho$. This weight function will be used to construct the Rodrigues formula for these polynomials which reads:

$$y_n(x) = \frac{B_n}{\rho(x)}\frac{d^n}{dx^n}\left[\sigma^n(x)\rho(x)\right], \qquad (3.6)$$

where $B_n$ is just a constant obtained by the normalization conditions, and $n = 0, 1, 2, \ldots$

## 4. The Solutions of the D-dimensional Radial equation

We use the NU method to solve equation (2.4), in the presence of our potential,

$$\frac{d^2 g(r)}{dr^2} + \frac{2\mu}{\hbar^2}\left[E - A\tanh^2(\lambda r) - \frac{B}{\tanh^2(\lambda r)} - \frac{\gamma_D \hbar^2}{2\mu}\frac{1}{r^2}\right]g(r) = 0, \qquad (4.1)$$

Where, $\gamma_D = \frac{(D+2l-2)^2 - 1}{4}$. Equation (4.1) cannot be solved analytically due to the centrifugal term $\frac{1}{r^2}$. Different authors used different approximation techniques to allow an approximate analytical solution of (4.1) and these methods rely on Taylor expansion of the centrifugal potential in terms of the other components of the potential of interest [39]. In this work, we use the following approximation obtained by Taylor expansion [40],

$$\frac{1}{\lambda^2 r^2} \approx -\frac{2}{3} - \frac{1}{3}\tanh^2(\lambda r) + \frac{1}{\tanh^2(\lambda r)}, \qquad (4.2)$$

The advantage of this approximation it is valid not only for $\lambda r \ll 1$ but also for $0 \leq \lambda r \leq 2$ with high accuracy. Also, it satisfies the limits on $\frac{1}{r^2}$ at zero and infinity, that is $\lim_{r \to 0} RHS = \frac{1}{\lambda^2 r^2}$ and $\lim_{r \to \infty} RHS = 0$, where *RHS* denotes the right-hand-side of (4.2). Now, Using Eq. (4.2) back in (4.1), we get

$$\left[\frac{d^2}{dr^2} - 4\lambda^2 \tilde{A} \tanh^2 \lambda r - \frac{4\lambda^2 \tilde{B}}{\tanh^2 \lambda r} + 4\lambda^2 \tilde{E}\right] g(r) = 0, \qquad (4.3)$$

where $4\lambda^2 \tilde{E} = \frac{2\mu}{\hbar^2}\left[E + \frac{\gamma_D \hbar^2 \lambda^2}{3\mu}\right]$, $4\lambda^2 \tilde{A} = \frac{2\mu}{\hbar^2}\left(A - \frac{\gamma_D \hbar^2 \lambda^2}{6\mu}\right)$, and $4\lambda^2 \tilde{B} = \frac{2\mu}{\hbar^2}\left[B + \frac{\gamma_D \hbar^2 \lambda^2}{2\mu}\right]$. Making change of variable $s = \tanh^2 \lambda r$ and by writing $g(s) = \phi(s) y(s)$, this transforms (4.3) to (3.1) with the polynomials being $\sigma = s - s^2$, $\tilde{\tau} = -(3s-1)/2$, and $\tilde{\sigma} = \tilde{E}s - \tilde{A}s^2 - \tilde{B}$. We now use Eq. (3.3.e) to calculate $\pi(s)$ which reads

$$\pi(s) = \frac{1-s}{4} \pm \sqrt{\left(\frac{1-s}{4}\right)^2 + \tilde{A}s^2 - \tilde{E}s + \tilde{B} + k(s - s^2)}, \qquad (4.4)$$

The choice of *k* that makes (4.4) a polynomial of first degree must satisfy $c_1^2 = c_2 c_3$, where $c_1 = 16k - 8 - 16\tilde{E}$, $c_2 = 1 + 16\tilde{A} - 16k$ and $c_3 = 4(1 + 16\tilde{B})$. This gives

$$\pi(s) = \frac{1-s}{4} \pm \frac{1}{4}\sqrt{c_2}\left(s + \frac{c_1}{2c_2}\right), \qquad (4.5)$$

where we will pick the negative part in (4.5) that makes $\tau' < 0$. The function $\tau(s)$ can be easily calculated using (3.3.b)

$$\tau(s) = 1 - 2s - \frac{1}{2}\sqrt{c_2}\left(s + \frac{c_1}{2c_2}\right), \qquad (4.6)$$

Using (4.5) and (4.6) in Eq. (3.5) and Eq. (3.3.f), we get

$$\left[\frac{4k}{(2n+1)} - (2n+1)\right]^2 = (1 + 16\tilde{A} - 16k), \qquad (4.7)$$

Solutions of (4.7) for *k* are

$$4k = -(2n+1)^2 \pm (2n+1)\sqrt{1+16\tilde{A}}, \tag{4.8}$$

The next step is to use the value of $c_2$ in Eq. (4.7) with the constraint on $k$ mentioned in Eq. (4.4), which is $c_1^2 = c_2 c_3$, we obtain

$$\left[8 + 16\tilde{E} - 16k\right]^2 = 4\left(1 + 16\tilde{B}\right)\left(1 + 16\tilde{A} - 16k\right), \tag{4.9}$$

The conditions for bound states are $k \leq 0$, and $\tilde{A}, \tilde{B} \geq -1/16$. Using (4.8) in (4.9), we write the bound states formula as follows:

$$\tilde{E}_{nl}^D = -\frac{1}{4}(2n+1)^2 - \frac{1}{4}(2n+1)\sqrt{1+16\tilde{A}} - \frac{1}{2}$$
$$\pm \frac{1}{8}\sqrt{\left(1+16\tilde{B}\right)\left(1+16\tilde{A}+4(2n+1)^2 + 4(2n+1)\sqrt{1+16\tilde{A}}\right)}, \tag{4.10}$$

In terms of the original parameters $A$, $B$, and $E$, the spectrum formula in $D$ dimensions' reads

$$\frac{2\mu}{\hbar^2} E_{nl}^D = -\frac{2\gamma_D \lambda^2}{3} - \lambda^2 (2n+1)^2$$
$$-\lambda^2 (2n+1)\sqrt{1 + \frac{8\mu}{\lambda^2 \hbar^2}\left(A - \frac{\gamma_D \hbar^2 \lambda^2}{6\mu}\right)} - 2\lambda^2 \pm \frac{\lambda^2}{2}\sqrt{\left(1 + \frac{8\mu}{\lambda^2 \hbar^2}\left[B + \frac{\gamma_D \hbar^2 \lambda^2}{2\mu}\right]\right)} \times \tag{4.11}$$
$$\sqrt{1 + \frac{8\mu}{\lambda^2 \hbar^2}\left(A - \frac{\gamma_D \hbar^2 \lambda^2}{6\mu}\right) + 4(2n+1)^2 + 4(2n+1)\sqrt{1 + \frac{8\mu}{\lambda^2 \hbar^2}\left(A - \frac{\gamma_D \hbar^2 \lambda^2}{6\mu}\right)}}$$

where conditions for bound states becomes $A \geq \frac{\lambda^2 \hbar^2}{2\mu}\left(\frac{\gamma_D}{3} - \frac{1}{4}\right)$, and $B \geq -\frac{\lambda^2 \hbar^2}{2\mu}\left(\gamma_D + \frac{1}{4}\right)$.

We will only take the (-) sign in (4.11) as explained below. The $s$-wave spectrum formula in three dimensions is the only exact solution which is obtained by setting $\gamma_D = 0$ in (4.11). However, for other higher states, the above solution is acceptable with high accuracy as far as the condition $0 \leq \lambda r \leq 2$ is satisfied.

The transformation $s = \tanh^2 \lambda x$ makes the domain of the function be $[0,1]$. This suggests a change of variable $z = 2s - 1$, to bring the domain to that of Jacobi polynomials which are well known classical orthogonal polynomials. By using (4.4) in Eq. (3.3.a), we obtain $\phi(z) = 2^{c_4 - 2c_5}(1+z)^{c_5}(1-z)^{c_5 - c_4}$, where $4c_4 = 1 + \sqrt{c_2}$, and $8c_5 = 2 - c_1/\sqrt{c_2}$. The weight function can be easily calculated using Eq. (4.6) in $(\sigma\rho)' = \tau\rho$, which gives $\rho(z) = 2^{2c_7 + c_6}(1+z)^{-c_7}(1-z)^{-c_6 - c_7}$, where $2c_6 = 8 + \sqrt{c_2}$ and $4c_7 = c_1 / \sqrt{c_2}$. The solution of Eq. (3.4) in our case is written in the following Rodrigues formula:

$$y_n(z) = C_n (1+z)^{c_7} (1-z)^{c_6+c_7} \frac{d^n}{dz^n}\left[(1+z)^{-c_7+n}(1-z)^{n-c_6-c_7}\right], \qquad (4.12)$$

By comparison to Jacobi polynomials, we conclude that $y_n(z) = P_n^{(-c_6-c_7,-c_7)}(z)$, where $P_n^{(-c_6-c_7,-c_7)}(z)$ is the Jacobi polynomial of order $n$ in $z$. Thus, the bound state solution of the radial wave equation now reads

$$g_n(r) = \Omega_n (\tanh \lambda r)^{-c_6-c_7+1/2} (\operatorname{sech} \lambda r)^{-c_7} P_n^{(-c_6-c_7,-c_7)}\left[2\tanh(\lambda r)^2 - 1\right], \qquad (4.13)$$

where $\Omega_n$ is just a normalization constant. We must clarify here that for Jacobi polynomials, we have to have $(-c_6-c_7), -c_7 > -1$. Thus, the parameters $c_1$ and $c_2$ are chosen to satisfy $c_1 < 4\sqrt{c_2}$, and $2c_2 + c_1 < -12\sqrt{c_2}$. Moreover, since those parameters depend on the energy as we mentioned previously, this yields us to reject the (+) sign in (4.10) and (4.11). Consequently, bound states occurs for $\tilde{B} > 35/16$ (which does not violate the old restriction $\tilde{B} \geq -1/16$), and $|\tilde{E}| > |k-1/2|$. The latter condition on $E$ is already satisfied as we can see in (4.11), so we don't have to worry about it.

To calculate the normalization constant $\Omega_n$, we first use the following identity of Jacobi polynomials [41]

$$P_n^{(a,b)}(y) = \frac{1}{2^n} \sum_{m=0}^{n} \binom{n+a}{m}\binom{n+b}{n-m} (1-y)^{m-n}(1+y)^m, \qquad (4.14)$$

Next, we use the normalization constraint $\int_0^\infty |g(r)|^2 dr = \int_{-1}^{+1} |g(y)|^2 \frac{dy}{\sqrt{2}\lambda\sqrt{1+y}(1-y)} = 1$, where $y = 2\tanh^2(\lambda r) - 1$, this gives

$$\frac{|\Omega_n|^2}{\lambda 2^{n+1/2}} \sum_{m=0}^{n} \binom{n+a}{m}\binom{n+b}{n-m} \int_{-1}^{+1} (1-y)^{2a+m-n-1}(1+y)^{b+m} P_n^{(a,b)}(y) dy = 1, \qquad (4.13)$$

To calculate the integral in (4.13), we will use the following very useful integral formula [42]

$$\int_{-1}^{+1} (1-y)^c (1+y)^d P_n^{(a,b)}(y) dy = \frac{2^{c+d+1} \Gamma(c+1)\Gamma(d+1)\Gamma(n+a+1)}{\Gamma(n+1)\Gamma(c+d+2)\Gamma(a+1)} \\ \times {}_3F_2(-n, n+a+b+1, c+1; a+1, c+d+2; 1)$$ (4.14)

where ${}_3F_2(a,b,c;d,e;f)$ is the generalized hypergeometric function [42]. By direct comparison between (4.13) and (4.14) we get $\Omega_n = 1/\sqrt{\Lambda_n}$, where $\Lambda_n$ is given below

$$\Lambda_n = \frac{1}{\lambda 2^{n+1/2}} \sum_{m=0}^{n} \binom{n+a}{m}\binom{n+b}{n-m} \frac{2^{2a+2m-n+b} \Gamma(2a+m-n)\Gamma(b+m+1)\Gamma(n+a+1)}{\Gamma(n+1)\Gamma(2a+2m-n+b+1)\Gamma(a+1)} \\ \times {}_3F_2(-n, n+a+b+1, 2a+m-n; a+1, 2a+2m-n+b+1; 1)$$ (4.15)

where $a = -c_6 - c_7$ and $b = -c_7$.

The only issue that is left for discussion in this section is that the solutions of the radial wave equation $\{g_n(r)\}_{n=0}^{N}$, where $N$ denotes the maximum number in which we get bound states, are not orthogonal! But they are normalized as we discussed above. We know that Hermitian operators with distinct eigenvalues must have orthogonal eigenvectors [43]. To solve this problem one must use the method of Gram-Schmidt (GS) to obtain an orthonormal set $\{\phi_n(r)\}_{n=0}^{N}$ by linear combinations [44]. The latter set will be the solutions of the radial wave equation. The process is a bit lengthy and we will not be able to do it here. However, we encourage the interested reader to do these calculations by referring to the process of GS.

## 5 Solutions of the angular equations

It is well-known from literature that solutions of (7) are written in terms of Jacobi polynomials as follows [42]

$$H(y) = N_n (1-y)^\alpha (1+y)^\beta P_n^{(c,d)}(y),$$ (5.1)

where $N_n$ is just a constant factor, $2\beta + j/2 = d+1$, and $2\alpha + j/2 = c+1$. Moreover, the latter parameters are written in terms of the quantum numbers as $c = d = c_j = l_{j-1} + (j-2)/2$, which yields $\alpha = \beta = l_{j-1}/2$ and $n = l_j - l_{j-1}$. The above solution was obtain using different methods including the NU technique [25]. Hence, solutions of (7) are written below

$$H(\theta_j) = N_n (\sin\theta_j)^{l_{j-1}} P_n^{(c_j, c_j)}(\cos\theta_j),$$ (5.2)

To solve Eq. (8), we introduce coordinate transformation as $y = \cos\theta_{D-1}$, which gives,

$$\left[(1-y^2)\frac{d^2}{dy^2}-(D-1)y\frac{d}{dy}+l(l+D-2)-\frac{\Lambda_{D-2}}{1-y^2}+U(y)\right]H(y)=0, \quad (5.3)$$

where $U(y)=\frac{2\mu}{\hbar^2}V_2(y)=(\gamma'y^2+\zeta'y+\kappa')/(1-y^2)$ for real parameters $\{\gamma',\zeta',\kappa'\}$ are related to $\{\gamma,\zeta,\kappa\}$ by a factor of $2\mu/\hbar^2$. Eq. (5.3) is of Hypergeometric type with the polynomials being $\sigma(y)=1-y^2$, $\tilde{\tau}(y)=-(D-1)y$, and $\tilde{\sigma}(y)=\eta_2 y^2+\eta_1 y+\eta_0$, where $\eta_2=\gamma'-l(l+D-2)$, $\eta_1=\zeta'$, and $\eta_0=\kappa'+l(l+D-2)-\Lambda_{D-2}$. The solutions of (5.3) are written as $H(y)=\phi(y)Y(y)$, where $\phi(y)$ satisfies (3.3.a). Next, we need to find the function $\pi(y)$ using (3.3.e), we find that this function takes the following form

$$\pi(y)=\frac{(D-3-2u_0)y}{2}+\frac{\eta_1}{2u_0}, \quad (5.4)$$

where $u_0=\sqrt{\left(\frac{D-3}{2}\right)^2-\eta_2-k}$, and the parameter $k$ defined in (3.3.f) must satisfy $4k-4\eta_0=\eta_1^2/u_0^2$. The latter constraint will be used later to obtain the eigenvalues of Eq. (5.3). Now, we use (5.4) in (3.3.b), we get $\tau(y)=-2(1+u_0)y+\frac{\eta_1}{u_0}$, which satisfies $d\tau(y)/dy<0$. Using (3.3.a), we can obtain $\phi(y)$ to be $\phi(y)=(1-y)^{-\frac{u_1+u_2}{2}}(1+y)^{\frac{u_2-u_1}{2}}$, where $2u_1=(D-3-2u_0)$, and $\eta_1=2u_0 u_2$. we can also calculate the weight function by solving $(\sigma\rho)'=\tau\rho$, which gives $\rho(y)=(1-y)^{u_0-u_2}(1+y)^{u_0+u_2}$. The Rodrigues formula of the polynomials $Y(y)$ reads

$$Y_n(y)=\frac{\xi_n}{(1-y)^{u_0-u_2}(1+y)^{u_0+u_2}}\frac{d^n}{dy^n}\left[(1-y)^{n+u_0-u_2}(1+y)^{n+u_0+u_2}\right], \quad (5.5)$$

where $\xi_n$ is just a constant. By direct comparison with the Rodrigues formula of Jacobi polynomials [xx], we conclude that $Y_n(y)=P_n^{(u_0-u_2,u_0+u_2)}(y)$. As required by Jacobi polynomials, we must impose that $u_0\pm u_2>-1$. Now, we use (5.4) and (3.5) in (3.3.f) to obtain the following quadratic formula for $k$

$$\left[k-n^2-n+\frac{(D-3)}{2}\right]^2=(2n+1)^2\left[\left(\frac{D-3}{2}\right)^2-\eta_2-k\right], \quad (5.6)$$

The solutions of (5.6) are given below

$$k=\frac{1}{2}\left[2-D-2n-2n^2\pm(1+2n)\sqrt{(D-2)^2-4\eta_2}\right], \quad (5.7)$$

Moreover, we use $4k - 4\eta_0 = \eta_1^2 / \left(\left(\dfrac{D-3}{2}\right)^2 - \eta_2 - k\right)$ to obtain another solution for $k$

$$8k = 9 + D^2 - 6D + 4\eta_0 - 4\eta_2 \pm \sqrt{((D-3)^2 + 4\eta_0 - 4\eta_2)^2 - 16(\eta_1^2 + \eta_0((D-3)^2 - 4\eta_2))}, \qquad (5.8)$$

Direct comparison between (5.7) and (5.8) gives

$$(2n+1)^2 = -\dfrac{(D-1)^2 + 4\eta_0 - 4\eta_2 - 2}{2}, \qquad (5.9)$$

$$(2n+1)^2 \left[(D-2)^2 - 4\eta_2\right] = \dfrac{((D-3)^2 + 4\eta_0 - 4\eta_2)^2}{16} - (\eta_1^2 + \eta_0((D-3)^2 - 4\eta_2)), \qquad (5.10)$$

In the next section, we will consider different examples and try to obtain the unknown parameters for each case.

## 6. Results and Discussions

In this section, we will discuss different examples that are considered as special cases of the potential in (1.1).

As a first example, we consider the case when $\gamma' = \zeta' = 0, \kappa' \neq 0$, which is equivalent to the following noncentral Hyperbolic potential

$$V(r,\theta) = A \tanh^2(\lambda r) + \dfrac{B}{\tanh^2(\lambda r)} + \dfrac{\kappa \csc^2 \theta}{r^2}, \qquad (6.1)$$

In this case, we have $u_2 = 0$. Thus, solutions of Eq. (5.3) reads

$$H_n(\theta_{D-1}) = N_n (\sin \theta_{D-1})^{-u_1} P_n^{(u_0, u_0)}(\cos \theta_{D-1}), \qquad (6.2)$$

where $2u_1 = (D-3-2u_0)$, $u_0 = \sqrt{\left(\dfrac{D-3}{2}\right)^2 + l(l+D-2) - k}$, and $k$ is given below

$$8k = 9 + D^2 - 6D + 4\eta_0 + 4l(l+D-2) \pm \sqrt{((D-3)^2 + 4\eta_0 + 4l(l+D-2))^2 - 16(\eta_0((D-3)^2 + 4l(l+D-2)))} \qquad (6.3)$$

and the corresponding eigenvalues is obtain from Eq.(5.10) as,

$$(2n+1)^2 \left[(D-2)^2 + 4l(l+D-2)\right] = \dfrac{((D-3)^2 + 4\eta_0 + 4l(l+D-2))^2}{16} \\ - (\eta_0((D-3)^2 + 4l(l+D-2))) \qquad (6.4)$$

(2)The next special case of our potential model is considered when we choose the ring shaped parameters $\gamma = \pm \varsigma, \kappa \neq 0$, which corresponds to the following potential

$$V(r,\theta) = A\tanh^2(\lambda r) + \frac{B}{\tanh^2(\lambda r)} + \frac{\gamma(\cot^2\theta \pm \cot\theta\csc\theta) + \kappa\csc^2\theta}{r^2} \quad (6.5)$$

Under these conditions, we have

$$\eta_1 = \varsigma, \eta_2 = \pm\varsigma - l(l+D-2), u_1 = \frac{(D-3-2u_0)}{2}, u_2 = \frac{\varsigma}{2u_0},$$

$$u_0 = \sqrt{\left(\frac{D-3}{2}\right)^2 \mp (\varsigma - l(l+D-2)) - k} \quad (6.6)$$

The $k$ values and the corresponding eigenvalues are obtained as follows:

$$8k = 9 + D^2 - 6D + 4\eta_0 \mp 4\varsigma + 4l(l+D-2) \pm$$
$$\sqrt{((D-3)^2 + 4\eta_0 \mp 4\varsigma + 4l(l+D-2))^2 - 16(\varsigma^2 + \eta_0((D-3)^2 \mp 4\varsigma + 4l(l+D-2)))} \quad (6.7)$$

$$(2n+1)^2\left[(D-2)^2 \mp 4\varsigma + 4l(l+D-2)\right] = \frac{((D-3)^2 + 4\eta_0 \mp 4\varsigma + 4l(l+D-2))^2}{16}$$
$$- (\varsigma^2 + \eta_0((D-3)^2 \mp 4\varsigma + 4l(l+D-2))) \quad (6.8)$$

The associated unnormalized wave function is obtain as,

$$H_n(\theta_{D-1}) = N_n(1-\cos\theta_{D-1})^{-\frac{u_1+u_2}{2}}(1+\cos\theta_{D-1})^{\frac{u_2-u_1}{2}} P_n^{(u_0-u_2,u_0+u_2)}(\cos\theta_{D-1}) \quad (6.9)$$

(3) Another special case of our study is when $\varsigma = 0, \gamma, \kappa \neq 0$, which corresponds to the following potential

$$V(r,\theta) = A\tanh^2(\lambda r) + \frac{B}{\tanh^2(\lambda r)} + \frac{(\gamma+\kappa)\cot^2\theta + \kappa}{r^2} \quad (6.10)$$

With these assumptions, we have
$$\eta_1 = 0, \eta_1 = \kappa + l(l+D-2) - \Lambda_{D-2}, \eta_2 = \gamma - l(l+D-2),$$
$$u_1 = \frac{D-3-2u_0}{2}, u_2 = 0, u_0 = \sqrt{\left(\frac{D-3}{2}\right)^2 - \gamma + l(l+D-2) + k} \quad (6.11)$$

Under this special case, we obtain the $k$ parameter, the eigenvalues and the corresponding wave function as follows:

$$8k = 9 + D^2 - 6D + 4(\kappa + l(l+D-2) - \Lambda_{D-2}) - 4(\gamma - l(l+D-2)) \pm$$
$$\sqrt{((D-3)^2 + 4(\kappa + l(l+D-2) - \Lambda_{D-2}) - 4(\gamma - l(l+D-2)))^2 - 16(\eta_0((D-3)^2 - 4\eta_2))} \quad (6.12)$$

,

$$(2n+1)^2\left[(D-2)^2 - 4(\gamma - l(l+D-2))\right] = \frac{((D-3)^2 + 4(\kappa + l(l+D-2) - \Lambda_{D-2}) - 4(\gamma - l(l+D-2)))^2}{16}$$
$$-((\kappa + l(l+D-2) - \Lambda_{D-2})((D-3)^2 - 4(\gamma - l(l+D-2)))) \quad (6.13)$$

,

$$H_n(\theta_{D-1}) = N_n (\sin\theta_{D-1})^{-u_1} P_n^{(u_0,u_0)}(\cos\theta_{D-1}) \quad (6.14)$$

However, one needs to be careful here as for $\gamma = -\kappa$ there will be no ring shaped term and one ends up with Hyperbolic PT potential plus pseudo centrifugal term

$$V(r,\theta) = A\tanh^2(\lambda r) + \frac{B}{\tanh^2(\lambda r)} + \frac{\kappa}{r^2} \quad (6.15)$$

(4) We consider the last special case for $\gamma = 0, \kappa = \pm\varsigma$, which corresponds to the following potential of the form,

$$V(r,\theta) = A\tanh^2(\lambda r) + \frac{B}{\tanh^2(\lambda r)} + \frac{\varsigma\left[\cot\theta\csc\theta \pm \csc^2\theta\right]}{r^2} \quad (6.16)$$

The following parameters are obtain under this case,
$\eta_1 = \pm\varsigma, \eta_2 = -l(l+D-2), \eta_0 = \pm\varsigma + l(l+D-2) - \Lambda_{D-2}$,

$$u_1 = \left(\frac{D-3-2u_0}{2}\right), u_0 = \sqrt{\left(\frac{D-3}{2}\right)^2 + l(l+D-2) - k}, u_2 = \frac{\pm\varsigma}{2u_0} \quad (6.17)$$

Using Eq.(6.17), we obtain the $k$-paramter, the eigenvalues and the corresponding wave function for this special case as follows:

$$8k = 9 + D^2 - 6D + 4(\pm\varsigma + l(l+D-2) - \Lambda_{D-2}) + 4l(l+D-2) \pm$$
$$\sqrt{((D-3)^2 + 4(\pm\varsigma + l(l+D-2) - \Lambda_{D-2}) + 4l(l+D-2))^2 - 16((\pm\varsigma)^2} \quad (6.18)$$
$$\overline{+(\pm\varsigma + l(l+D-2) - \Lambda_{D-2})((D-3)^2 + 4l(l+D-2)))}$$

$$(2n+1)^2\left[(D-2)^2 + 4l(l+D-2)\right] = \frac{((D-3)^2 + 4(\pm\varsigma + l(l+D-2) - \Lambda_{D-2}) + 4l(l+D-2))^2}{16}$$
$$-((\pm\varsigma)^2 + (\pm\varsigma + l(l+D-2) - \Lambda_{D-2})((D-3)^2 + 4l(l+D-2))) \quad (6.19)$$

$$H_n(\theta_{D-1}) = N_n (1-\cos\theta_{D-1})^{-\frac{u_1+u_2}{2}} (1+\cos\theta_{D-1})^{\frac{u_2-u_1}{2}} P_n^{(u_0-u_2,u_0+u_2)}(\cos\theta_{D-1}) \quad (6.20)$$

## 7. Conclusions

In this paper, we have obtained analytically the solutions of the $D$-dimensional Schrödinger potential with hyperbolic Poschl Teller potential plus a generalized ring-shaped term. We employed NU and trial function methods to solve the radial and angular part of the Schrödinger equation respectively. This result is new and has never been reported before in the available literature to the best of our knowledge. Finally, this result can find many applications in atomic and molecular physics.

## Appendix A: Jacobi Polynomials

Jacobi polynomials $P_n^{(\mu,\nu)}(y)$ defined on $[-1, 1]$ are solutions of the following second order linear differential equation [8]:

$$\left\{(1-y^2)\frac{d^2}{dy^2}-[(\mu+\nu+2)y+\mu-\nu]\frac{d}{dy}+n(n+\mu+\nu+1)\right\}P_n^{(\mu,\nu)}(y)=0, \quad (A1)$$

We also mention their Orthogonality relation:

$$\int_{-1}^{1}(1-y)^\mu(1+y)^\nu P_n^{(\mu,\nu)}P_m^{(\mu,\nu)}dy=\frac{2^{\mu+\nu+1}}{2n+\mu+\nu+1}\frac{\Gamma(n+\mu+1)\Gamma(n+\nu+1)}{\Gamma(n+\mu+\nu+1)n!}\delta_{n,m}, \quad (A2)$$

## Appendix B: Hyperspherical Coordinates

The $D$-dimensional position vector $\vec{x}=(r,\theta_1,....\theta_{D-1})$ is defined in terms of Hyperspherical Cartesian coordinates below [37]:

$$x_1 = r\cos\theta_1 \sin\theta_2...\sin\theta_{D-1}, \quad (B1)$$
$$x_2 = r\sin\theta_1 \sin\theta_2...\sin\theta_{D-1}, \quad (B2)$$
$$x_j = r\cos\theta_{j-1} \sin\theta_j...\sin\theta_{D-1}, \quad (B3)$$

where $j=3,4,....,D-1$, $x_D = r\cos\theta_{D-1}$, and $\sum_{j=1}^{D}x_j^2 = r^2$. For $D = 2$, this is the case of polar coordinates $(r,\varphi)$ with $x_1 = x = r\cos\varphi$ and $x_2 = y = r\sin\varphi$, whereas $D = 3$ represents the spherical coordinates $(r,\varphi,\theta)$, where $x_1 = x = r\cos\varphi\sin\theta$, $x_2 = y = r\cos\varphi\sin\theta$, and $x_3 = z = r\cos\theta$.

The volume element in $D$-dimension is defined to be $dV = r^{D-1}dr\prod_{j=1}^{D-1}(\sin\theta_j)^{j-1}d\theta_j$, where $r \in [0,\infty[$, $\theta_1 \in [0,2\pi]$, and $\theta_j \in [0,\pi]$ for $j \geq 2$. The Laplacian operator in $D$ dimensions is defined below

$$\nabla_D^2 = \frac{\partial^2}{\partial r^2} + \frac{D-1}{r}\frac{\partial}{\partial r} + \frac{1}{r^2} \times \left[\frac{1}{\sin^{D-2}\theta_{D-1}}\frac{\partial}{\partial \theta_{D-1}}\left(\sin^{D-2}\theta_{D-1}\frac{\partial}{\partial \theta_{D-1}}\right) - \frac{L_{D-2}^2}{\sin^2\theta_{D-1}}\right], \quad (B4)$$

Finally, we mention the normalization conditions of the wavefunction in the $D$ dimensions

$$\int_0^\infty |g_n(r)|^2 dr = 1, \quad (B5)$$

$$\prod_{j=2}^{D-1}\int_0^\pi |H(\theta_j)|^2 (\sin\theta_j)^{j-1}d\theta_j = 1, \quad (B6)$$